\documentclass[prx,twocolumn,superscriptaddress]{revtex4-1}

\usepackage{amsmath}
\usepackage{amssymb}
\usepackage{amsthm}
\usepackage{amsfonts}
\usepackage{listings}
\usepackage{physics}
\usepackage{enumerate}
\usepackage{latexsym}
\usepackage{color}
\usepackage{bm}
\usepackage{hyperref}
\hypersetup{
 pdfnewwindow=true, colorlinks=true,
 linkcolor=blue, anchorcolor=blue,
 citecolor=blue, filecolor=blue,
 menucolor=blue, urlcolor=blue}

\usepackage{psfrag}

\usepackage{bm}
\usepackage{graphicx}
\usepackage{subfigure}

\usepackage{multirow}
\usepackage{array}
\usepackage{xcolor}

\RequirePackage[normalem]{ulem} 
\RequirePackage{color}\definecolor{RED}{rgb}{1,0,0}\definecolor{BLUE}{rgb}{0,0,1}

\input{epsf}

\newcommand{\nc}{\newcommand}
\nc{\webirvsp}{\href{https://github.com/zjwang11/irvsp}{\texttt{IRVSP}} }
\nc{\webirtb}{\href{https://github.com/zjwang11/irvsp}{\texttt{ir2tb}} }
\nc{\webirpw}{\href{https://github.com/zjwang11/ir2pw}{\texttt{ir2ph}} }
\nc{\webchecktopmat}{\href{https://www.cryst.ehu.es/cryst/checktopologicalmagmat}{\texttt{Check Topological Mat}}}
\nc{\webposabr}{\href{https://github.com/zjwang11/UnconvMat/blob/master/src_pos2aBR.tar.gz}{\texttt{POS2ABR}} }
\nc{\webUnconvMat}{\href{http://tm.iphy.ac.cn/UnconvMat.html}{\texttt{UnconvMat}} }
\nc{\online}{\href{http://tm.iphy.ac.cn/UnconvMat.html}{online}}

\begin{document}

%\linenumbers

\tolerance 10000

\newcommand{\vk}{{\bf k}}

\draft

\title{Spin-polarized triplet excitonic insulators in Ta$_3X_8$ ($X$=I, Br) monolayers}

% in English titles articles and words like to, on, at etc are always spelled with small letters
\author{Haohao Sheng}
%\thanks{These authors contributed equally to this work.}
\affiliation{Beijing National Laboratory for Condensed Matter Physics,
and Institute of Physics, Chinese Academy of Sciences, Beijing 100190, China}
\affiliation{University of Chinese Academy of Sciences, Beijing 100049, China}

\author{Jingyu Yao}
\affiliation{Beijing National Laboratory for Condensed Matter Physics,
and Institute of Physics, Chinese Academy of Sciences, Beijing 100190, China}
\affiliation{University of Chinese Academy of Sciences, Beijing 100049, China}

\author{Sheng Zhang}
\affiliation{Beijing National Laboratory for Condensed Matter Physics,
and Institute of Physics, Chinese Academy of Sciences, Beijing 100190, China}
\affiliation{University of Chinese Academy of Sciences, Beijing 100049, China}

\author{Quansheng Wu}
\affiliation{Beijing National Laboratory for Condensed Matter Physics,
and Institute of Physics, Chinese Academy of Sciences, Beijing 100190, China}
\affiliation{University of Chinese Academy of Sciences, Beijing 100049, China}

\author{Zhong Fang}
\affiliation{Beijing National Laboratory for Condensed Matter Physics,
and Institute of Physics, Chinese Academy of Sciences, Beijing 100190, China}
\affiliation{University of Chinese Academy of Sciences, Beijing 100049, China}

\author{Xi Dai}
\affiliation{Department of Physics, Hong Kong University of Science and Technology, Hong Kong 999077, China}

\author{Hongming Weng}
\affiliation{Beijing National Laboratory for Condensed Matter Physics,
and Institute of Physics, Chinese Academy of Sciences, Beijing 100190, China}
\affiliation{University of Chinese Academy of Sciences, Beijing 100049, China}

\author{Zhijun Wang}
\email{wzj@iphy.ac.cn}
\affiliation{Beijing National Laboratory for Condensed Matter Physics,
and Institute of Physics, Chinese Academy of Sciences, Beijing 100190, China}
\affiliation{University of Chinese Academy of Sciences, Beijing 100049, China}

\begin{abstract}
Bose-Einstein condensation of spin-polarized triplet excitons can give rise to an intriguing spin supercurrent, enabling experimental detection of exciton condensation.
In this work, we predict that Ta$_3X_8$ ($X$=I, Br) ferromagnetic monolayers are spin-polarized triplet excitonic insulators (EIs), based on the systematic first-principles $GW$ calculations coupled with the Bethe-Salpeter equation ($GW$+BSE).
The single-particle calculations of spin-polarized band structures reveal that these monolayers are bipolar magnetic semiconductors, where the highest valence band and the lowest conduction band possess opposite spin polarization.
The two low-energy bands, primarily originating from Ta $d_{z^2}$ orbitals, are almost flat. 
The same-orbital parity and opposite-spin natures of the band-edge states effectively suppress dielectric screening, promoting the emergence of the EI state.
The $GW$+BSE calculations reveal that the binding energy of the lowest-energy exciton is 1.499 eV for Ta$_3$I$_8$ monolayer and 1.986 eV for Ta$_3$Br$_8$ monolayer. Since both values exceed the respective $GW$ band gaps, these results indicate a strong excitonic instability in these monolayers.
A wavefunction analysis confirms that the lowest-energy exciton is a tightly bound Frenkel-like state, exhibiting a spin-polarized triplet nature with $S_z=1$.
Our findings establish a valuable material platform for investigating spin-polarized triplet EIs, offering promising potential for spintronic applications.
\end{abstract}

\maketitle

\paragraph*{Introduction.}
Excitons are electron-hole pairs bound by attractive Coulomb interactions. In semiconductors or semimetals, when the exciton binding energy ($E_b$) exceeds the single-particle band gap ($E_g$), the spontaneous formation of excitons can cause a renormalization of the single-electron band structure~\cite{EI-first-PhysRev.158.462,EI-first-PhysRevLett.19.439,EI-first-RevModPhys.40.755,EI-first-doi:10.1080/14786436108243318}. 
This excitonic instability results in a novel many-body electronic state known as an excitonic insulator (EI). Characterized by the spontaneous Bose-Einstein condensation (BEC) of excitons~\cite{BEC-RevModPhys.42.1,BEC-10.1038/nature03081,BEC-doi:10.1126/science.aam6432}, the time reversal (TR) invariant EI behaves as a perfect insulator for both charge and spin transport.
In contrast, a spin-polarized triplet EI is referred to as a spin superconductor~\cite{spinsuperconductor-PhysRevB.84.214501}.
Although some materials have been theoretically predicted to be spin-polarized triplet EIs, such as ABC-stacked trilayer graphene~\cite{spinsuperconductor2-ABC-PhysRevB.86.085441},
ABCA-stacked tetralayer graphene~\cite{spinsuperconductor3-ABCA-PhysRevB.110.174512}, and semihydrogenated graphene~\cite{TripletFM-PhysRevLett.124.166401}, no experimental evidence has been reported yet.

To achieve an EI, it is essential to significantly reduce the screening of Coulomb interactions. 
Dimensionality reduction can weaken electron-hole screening and enhance their binding in two-dimensional (2D) systems~\cite{RevModPhys.90.021001,PhysRevLett.113.076802}. 
Further reduction of screening can be achieved by targeting band-edge states with opposite spin components~\cite{TripletFM-PhysRevLett.124.166401}, the same parity~\cite{parity1-PhysRevB.98.081408-2018,parity-PhysRevB.104.085133-2021,parity-tp-PhysRevB.107.235147-2023,parity-tp-triplet-PhysRevB.109.075167-2024,half-Excitonic-PhysRevLett.122.236402,M6Te6-acs.nanolett.4c05448}, and the same $C_{2z}$ symmetry eigenvalues~\cite{Yao_2024}.
To date, 2D EIs confirmed by both theory and experiment remain scarce, with reported candidates limited to the InAs/GaSb quantum-well~\cite{tpEI-Du2017-InAs/GaSb} and the monolayer of $1T'$-phase WTe$_2$~\cite{tpEI-Jia2022-WTe2,tpEI-Sun2022-WTe2}.
The Kagome lattice offers a valuable platform for exploring the interactions between geometry, topology, correlation, magnetism, and more~\cite{kagome-Ghimire2020,kagome-Teng2022,kagome-Yin2022,djz-PhysRevB.108.115123}. 
In particular, niobium halide clusters are noteworthy due to their breathing Kagome geometry~\cite{Nb3X8-jacs}, which supports flat band~\cite{Nb3X8-flat-Regmi2022,Nb3X8-flat-PhysRevB.108.L121404,Nb3Cl8-flat-nanolett.2c00778,Nb3X8-flat-PhysRevMaterials.5.084203,Nb3X8-mott-PhysRevX.13.041049,Nb3X8-mott-PhysRevB.107.035126}.   
This material family features weak interlayer van der Waals (vdW) interactions, allowing for straightforward thinning down to a 2D limit through mechanical exfoliation~\cite{Nb3X8-2D-OH2020154877,Nb3X8-2D-RRL}.
Recently, Ta$_3X_8$ ($X$=I, Br) monolayers are predicted to be stable 2D intrinsic multiferroic semiconductors with the coexistence of ferromagnetic (FM), ferroelectric, and ferrovalley orders~\cite{Ta3X8-acsaelm.2c00464,Ta3I8-acs.jpclett.4c00858,Ta3I8-APL}.

\begin{figure}[!htb]
\centering
\includegraphics[width=8.5 cm]{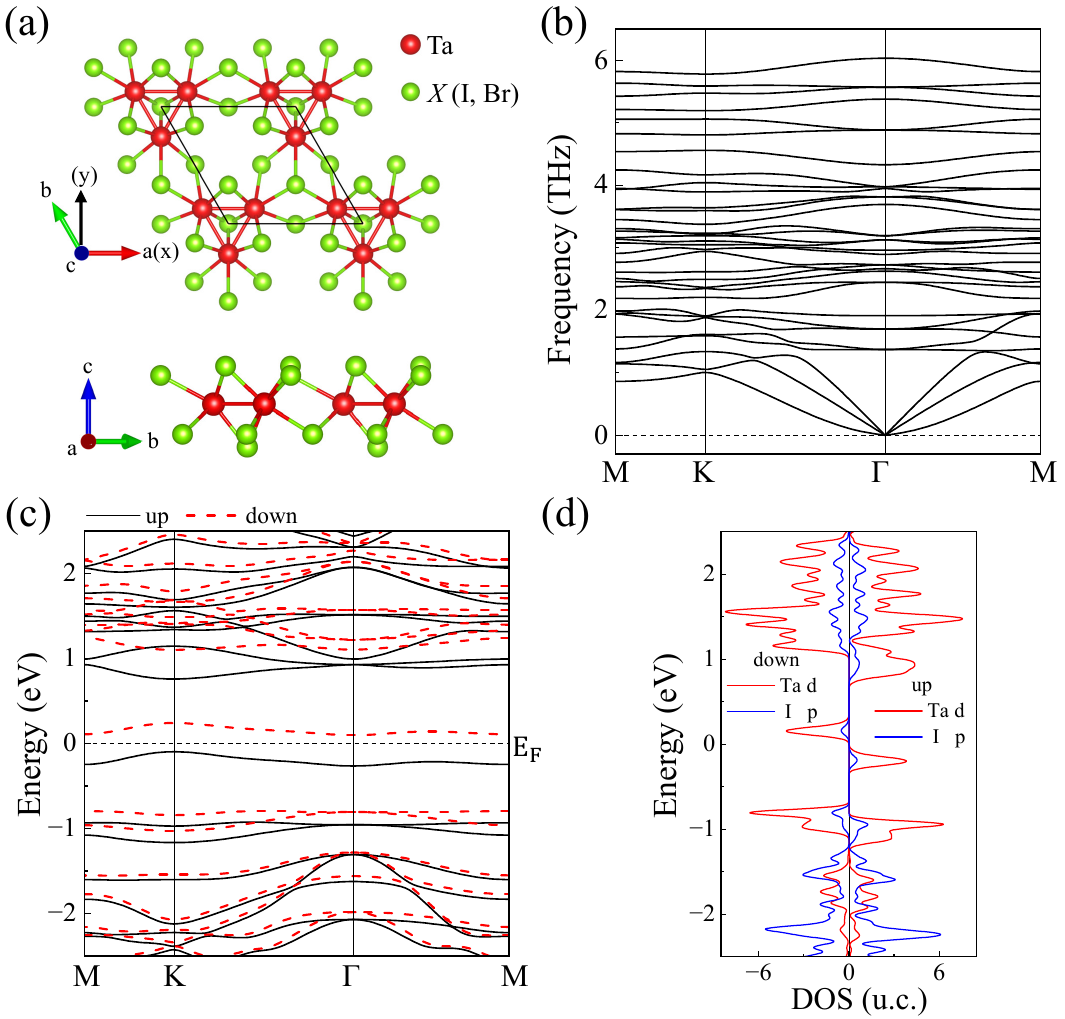}
\caption{(Color online) The (a) crystal structure, (b) phonon spectra, and (c,d) electronic band structures of Ta$_3$I$_8$ monolayer. 
The unit cell is outlined in black in panel (a). 
(c) The spin-polarized band structure and (d) partial densities of states (DOS) without spin-orbit coupling (SOC).
Black and red dotted lines represent the spin-up and spin-down bands, respectively.
} \label{fig-TaI-band}
\end{figure}

In this work, we predict that Ta$_3X_8$ monolayers are spin-polarized triplet EIs, as demonstrated by systematic first-principles $GW$ calculations coupled with Bethe-Salpeter equation ($GW$+BSE).
The single-particle calculations reveal that these monolayers are bipolar magnetic semiconductors (BMSs), where the highest valence band (VB) and the lowest conduction band (CB) are fully spin-polarized in opposite spin directions.
Moreover, the two low-energy bands from the Ta $d_{z^2}$ orbitals exhibit minimal energy dispersion, forming flat bands.
The characteristics of same-orbital parity and opposite-spin band-edge states effectively suppress band-edge transitions and dielectric screening, facilitating the realization of the EI state. 
The $GW$+BSE calculations show that the $GW$ band gap is 1.331 (resp. 1.722) eV, while the $E_b$ of the lowest-energy exciton reaches 1.499 (resp. 1.986) eV for Ta$_3$I$_8$ (resp. Ta$_3$Br$_8$) monolayer, exceeding the corresponding $GW$ band gap.
A wavefunction analysis confirms that the lowest-energy exciton is a spin-polarized triplet state with $ S_z =1$. 
The results indicate that FM Ta$_3$I$_8$ and Ta$_3$Br$_8$ monolayers are spin-polarized triplet EIs, where the spontaneous exciton BEC can generate an intriguing spin supercurrent.

\begin{figure}[!htb]
\centering
\includegraphics[width=8.5 cm]{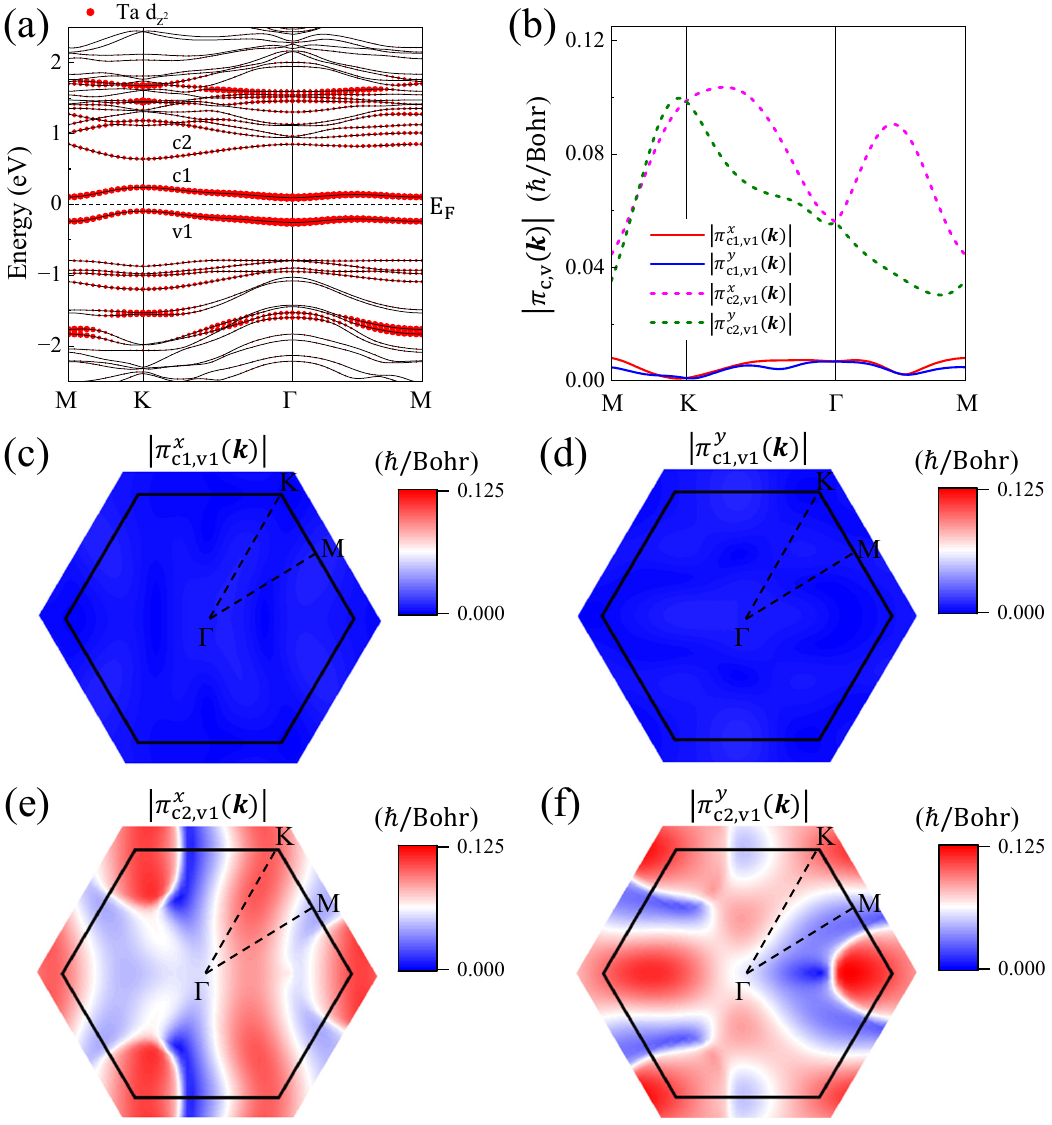}
\caption{(Color online) 
(a) The orbital-resolved band structure with SOC of Ta$_3$I$_8$ monolayer. The size of red dots represents the weight of the Ta $d_{z^2}$ orbitals.
(b) The modulus of the generalized momentum matrix elements 
$ \boldsymbol{\pi}_{\text{c1(c2)},\text{v1}}(\boldsymbol{k}) $
along the high-symmetry paths. Here, $\text{v1}$ and $\text{c1}$ ($\text{c2}$) refer to the highest valence band and the lowest (second-lowest) conduction band, respectively.
(c-f) The modulus of the $ \boldsymbol{\pi}_{\text{c1(c2)},\text{v1}}(\boldsymbol{k}) $ in the first Brillouin zone, including
(c) $ \pi_{\text{c1},\text{v1}}^x(\boldsymbol{k}) $, 
(d) $ \pi_{\text{c1},\text{v1}}^y(\boldsymbol{k}) $,
(e) $ \pi_{\text{c2},\text{v1}}^x(\boldsymbol{k}) $, and
(f) $ \pi_{\text{c2},\text{v1}}^y(\boldsymbol{k}) $.
} \label{fig-TaI-band2}
\end{figure}

\paragraph*{Crystal and electronic structure.} 
The Ta$_3X_8$ monolayers form a breathing Kagome lattice with space group $P3m1$, as illustrated in Fig.~\ref{fig-TaI-band}(a). 
The structure is derived by removing a Ta atom from the 2$\times$2 supercell of the $1T$-phase Ta$X_2$ and introducing a breathing distortion, resulting in Ta-trimer clusters in the Kagome lattice.
First-principles single-particle calculations show that they exhibit a FM ground state with a total magnetic moment of 1 $\mu_B$ per unit cell.
Details of the calculation methods are provided in Section \textcolor{blue}{A} of the Supplementary Material (SM).
Since Ta$_3X_8$ monolayers have the same properties, we take Ta$_3$I$_8$ monolayer as an example in the main text (see results for Ta$_3$Br$_8$ monolayer in the SM).
The phonon dispersion of the FM state of Ta$_3$I$_8$ monolayer is presented in Fig.~\ref{fig-TaI-band}(b), indicating that it is dynamically stable.
In the spin-polarized band structure shown in Fig.~\ref{fig-TaI-band}(c), the highest VB and the lowest CB exhibit opposite spin directions, a hallmark feature of BMSs~\cite{BMS-2012,BMS-10.1093/nsr/nww026-2016,BMS-2022-spin-filtering}.
The typical BMS band structure remains robust under the GGA+U method for Ta$_3X_8$, but not for Nb$_3X_8$ (see the SM).
By applying a gate voltage, it is possible to generate and manipulate completely spin-polarized currents with reversible spin polarization.
As a result, BMSs hold considerable promise for application in spintronic devices, such as bipolar field-effect spin filter~\cite{BMS-2012,BMS-2014-spin-filtering,BMS-2022-spin-filtering}.

Moreover, the two low-energy bands are almost flat (with $\sim 0.1$ eV bandwidth), suggesting strong interelectronic correlation within the systems.
Considering spin-orbit coupling (SOC), the band structure is shown in Fig.~\ref{fig-TaI-band2}(a), which demonstrates that SOC has little impact on the two flat bands. 
As evidenced by the partial densities of states and the orbital-resolved band structure in Fig.~\ref{fig-TaI-band}(d) and Fig.~\ref{fig-TaI-band2}(a), the low-energy flat bands are mainly from Ta $d_{z^2}$ orbitals.

\begin{figure}[!htb]
\centering
\includegraphics[width=8.5 cm]{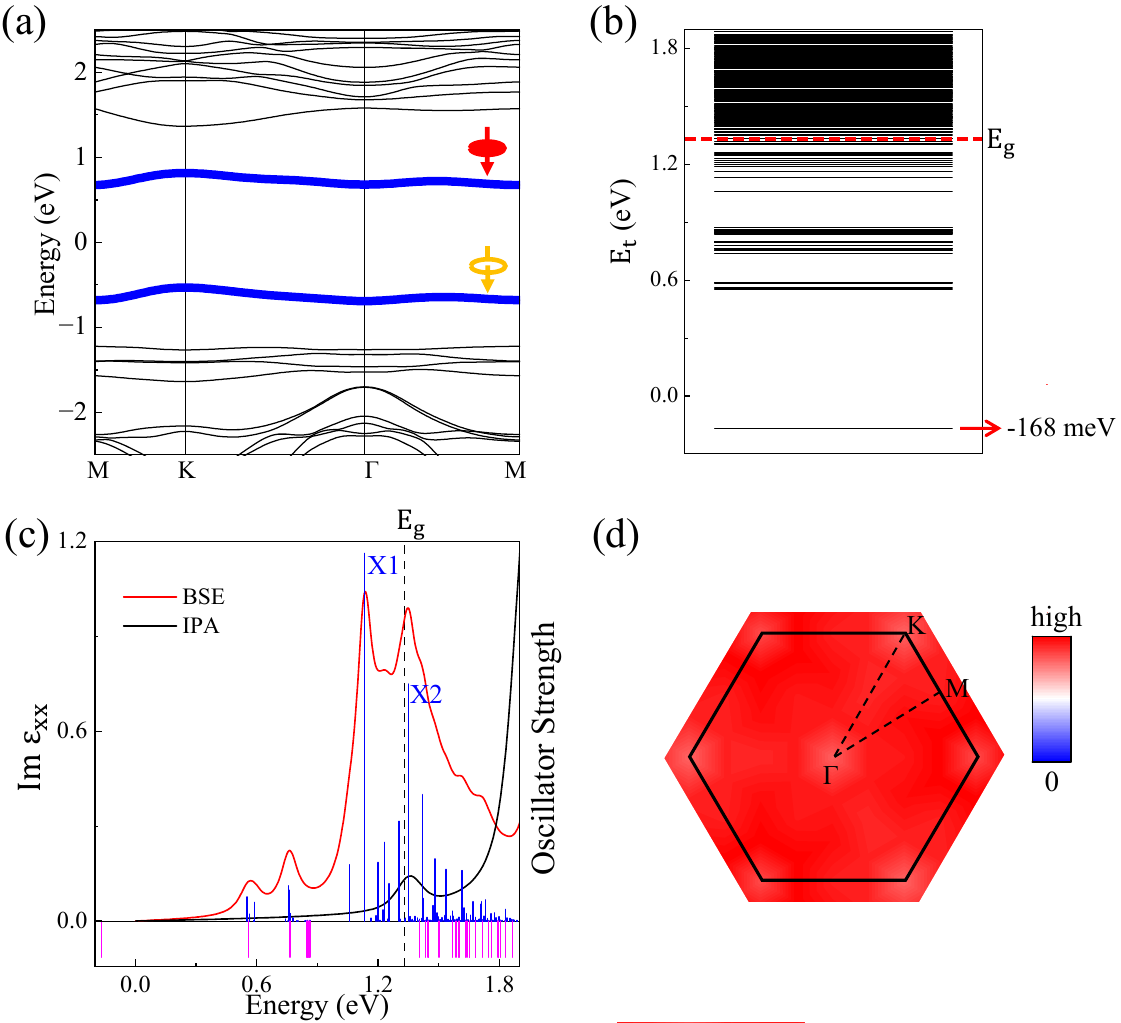}
\caption{(Color online) Low-energy excitons with $\mathbf{q}=0$ of Ta$_3$I$_8$ monolayer.
(a) The $G_0W_0$ band structure and the BSE fatband. The spread of color on the bands corresponds to the contribution of a particular electron-hole transition to the lowest-energy exciton. The orange arrow with a circle denotes the spin of the hole left behind after excitation and the red arrow denotes the spin of the electron. We only show the weights of transitions between the $\text{v1}$ and $\text{c1}$ bands, as contributions from other electron-hole pairs are more than 15 times smaller.
(b) Exciton transition energy ($E_t$) spectrum. 
Each horizontal line corresponds to an exciton state. The lowest-energy exciton exhibits a negative $E_t$. 
(c) Imaginary part of the dielectric function (left axis) and exciton oscillator strength (right axis). The red line is obtained from the BSE calculation, while the black line is obtained from the independent-particle approximation (IPA), i.e., ignoring electron-hole interactions. 
The bright excitons are depicted by blue vertical lines, with the highness indicating the oscillator strength, while the dark excitons are shown by the pink vertical lines under the $x$-axis, whose strengths are less than $5 \times 10^{-6}$ of that of the brightest exciton.
(d) Exciton wavefunction in reciprocal space for the lowest-energy exciton. Its substantial delocalization in reciprocal space corresponds to a high localization in real space.
} 
\label{fig-TaI-EI}
\end{figure}

\begin{figure}[!htb]
\centering
\includegraphics[width=8.5 cm]{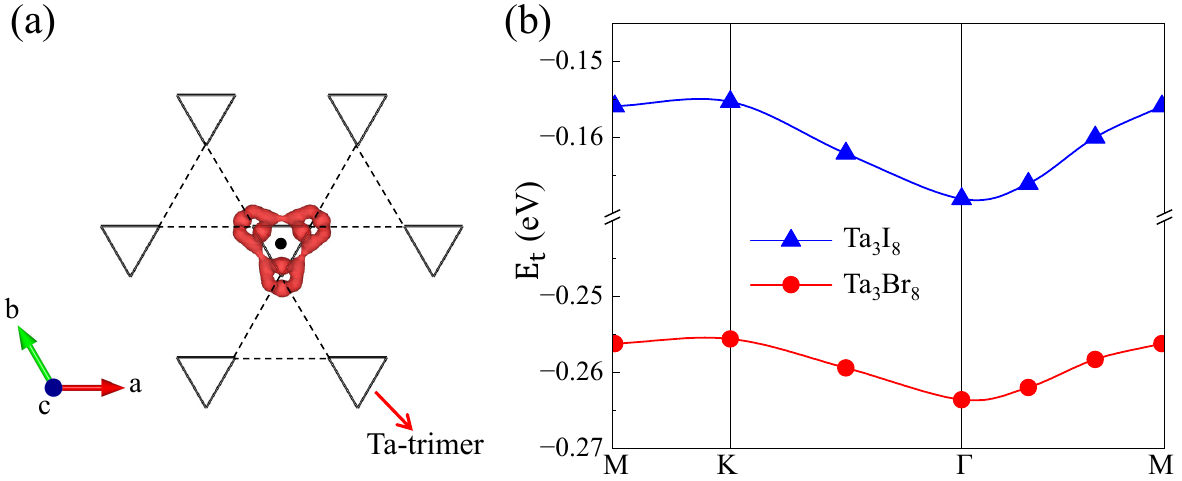}
\caption{(Color online) 
(a) Exciton wavefunction in real space of the lowest-energy exciton for Ta$_3$I$_8$ monolayer.
Only the Ta-breathing Kagome lattice and a portion of the 6$\times$6$\times$1 supercell are shown.
The contour plot (red) is the probability density of locating the bound electron once the hole position is fixed (black dot).
(b) Exciton dispersion of the lowest-energy exciton along the high-symmetry paths for Ta$_3X_8$ monolayers.
} \label{fig-EI-q}
\end{figure}

\paragraph*{Forbidden band-edge transition.} 
In 2D materials, a previous study has shown that $E_b$ often scales with $E_g$, typically as $E_b$ $\approx$ $E_g$/4~\cite{1/4-PhysRevLett.118.266401}.
Therefore, to achieve a 2D EI, one has to break this relationship in the materials, by reducing the dielectric constant (screening of Coulomb interactions).
For this purpose, we directly compute the dielectric constant $\varepsilon$ in the random phase approximation approach and in the absence of the local field effect~\cite{RPA-PhysRevB.73.045112},
\begin{equation}
\label{epsilon}
    \varepsilon_{\alpha\beta} 
    = \delta_{\alpha\beta}+\frac{8\pi e^2 \hbar^2}{m_e^2}   
    \sum_{c,v}\int_{\boldsymbol{k}\in BZ}
    \frac{ \pi_{cv}^\alpha(\boldsymbol{k})  
           \pi_{vc}^\beta(\boldsymbol{k})}{(E_{c,\boldsymbol{k}}-E_{v,\boldsymbol{k}})^3}\frac{d\boldsymbol{k}}{(2\pi)^3},
\end{equation}
where $\boldsymbol{\pi}_{cv}(\boldsymbol{k})$ is the generalized momentum matrix including SOC, given by
\begin{equation}
\label{phi}
\boldsymbol{\pi}_{cv}(\boldsymbol{k}) \equiv \langle u_{c,\boldsymbol{k}}\vert \hat{\vb* p}+\frac{1}{2mc^2}\left(\hat{{\vb* s}}\times\nabla V({\vb* r})\right) \vert u_{v,\boldsymbol{k}} \rangle .
\end{equation}
Here, $ \hat{\vb* p} $ is the momentum operator, $ V({\vb* r}) $ is the potential in crystal, $ \hat{{\vb* s}} $ is the spin momentum operator, $u_{c,\boldsymbol{k}}$ and $u_{v,\boldsymbol{k}}$ refer to the periodic parts of CB and VB Bloch states, respectively.
The $\boldsymbol{\pi}_{cv}(\boldsymbol{k})$ can be calculated by the VASP2KP package~\cite{vasp2kp}. 
Figs.~\ref{fig-TaI-band2}(c,d) show the results of $\boldsymbol{\pi}_{cv}(\boldsymbol{k})$ between the $\text{v1}$ and $\text{c1}$ bands in the first Brillouin zone (BZ).
For comparison, the results between the $\text{v1}$ and $\text{c2}$ bands are presented in Figs.~\ref{fig-TaI-band2}(e,f). The $\text{v1}$ and $\text{c1}$ ($\text{c2}$) are the highest VB and the lowest (second-lowest) CB, respectively. 
The $ \boldsymbol{\pi}_{\text{c1(c2)},\text{v1}}(\boldsymbol{k}) $ along the high-symmetry paths is also presented in Fig.~\ref{fig-TaI-band2}(b).
One can see that the $ \boldsymbol{\pi}_{\text{c1},\text{v1}}(\boldsymbol{k}) $ is extremely low, with modulus less than 0.01 $\hbar/\mathrm{Bohr}$.
This result implies that the two low-energy flat bands do not contribute significantly to the dielectric constant, leading to an unusually low 2D polarization.  
These results can decouple $E_b$ from $E_g$ and significantly enhance $E_b$, thereby enabling the realization of the EI state in the nanolayer.

We attribute the minor values of $\boldsymbol{\pi}_{\text{c1},\text{v1}}(\boldsymbol{k})$ to the following reasons. First, the low-energy flat Ta $d$ orbital bands suggest that Ta atoms maintain good localized atomic characteristics. According to the selection rules for atomic orbital transitions, the low-energy $d$-$d$ transitions are parity-forbidden~\cite{half-Excitonic-PhysRevLett.122.236402}. Second, in systems with weak SOC, electric-dipole transitions adhere to the spin selection rule~\cite{TripletFM-PhysRevLett.124.166401}. As a result, the transitions between the $\text{v1}$ and $\text{c1}$ bands are spin-forbidden.

\paragraph*{Exciton with direct transition.} 
To analyze low-energy excitons, we have performed $GW$+BSE calculations (Section \textcolor{blue}{B} of the SM).
As shown in Fig.~\ref{fig-TaI-EI}(a), a more accurate band structure is obtained by many-body single-shot $GW$ calculations ($G_0W_0$) at the PBE level, where the $E_g$ changes from 0.334 eV (PBE) to 1.331 eV ($GW$). 
Based on the $G_0W_0$ electronic structure, we solve the BSE using ten VBs and ten CBs.
Fig.~\ref{fig-TaI-EI}(b) shows the exciton transition energies ($E_t$) for all direct excitons ($\mathbf{q}=0$). 
The lowest-energy exciton exhibits a negative $E_t$ of -168 meV, indicating that the $E_b$ exceeds the $E_g$.
This implies that the Ta$_3$I$_8$ monolayer exhibits a many-body EI ground state characterized by the spontaneous exciton BEC. 
The large value of $\abs{E_t}$ indicates that the EI state can survive at room temperature.
The lowest-energy exciton is identified as a dark state due to its extremely low oscillator strength.
The imaginary part of the frequency-dependent dielectric function, obtained from the BSE calculations, is presented in Fig.~\ref{fig-TaI-EI}(c).
Compared to the independent-particle approximation (IPA), two prominent optical absorption peaks emerge, originating from two bright excitons (X1 and X2).
The X1 and X2 exciton states arise from optically allowed transitions between the 
$\text{v1}$ and $\text{c2}$ bands.

We primarily focus on the lowest-energy exciton state with a negative $E_t$.
Fig.~\ref{fig-TaI-EI}(d) illustrates the exciton reciprocal-space wavefunction, which extends across the entire BZ. 
Its substantial delocalization in reciprocal space corresponds to a high localization in real space, resembling a tightly bound Frenkel-like exciton.
As shown in Fig.~\ref{fig-EI-q}(a), when a hole is introduced at the center of the Ta-trimer, the bound electron becomes localized exclusively on the nearest Ta atoms.
The BSE fatband~\cite{BSEfatband-srep28618} is plotted in Fig.~\ref{fig-TaI-EI}(a). 
We observe that the lowest-energy exciton arises solely from the $d$-$d$ transitions with spin-flipping between two low-energy bands.
This spin-flipping in spin-polarized electron-hole transitions can produce spin-polarized triplet excitons with $ S_z =1$.

To ensure accuracy, we have also performed $GW$+BSE calculations at different levels, including PBE+U and HSE06. 
Different methods do not significantly change the band structure.
While the $E_g$ varies across different methods, the $E_t$ of the lowest-energy exciton remains consistently negative (Section \textcolor{blue}{C} of the SM). 
On the other hand, the spin-polarized triplet EI state remains robust across various directions of magnetic moments, as shown in Section \textcolor{blue}{D} of the SM.
Section \textcolor{blue}{E} of the SM presents the calculations of Ta$_3$Br$_8$ monolayer, which also exhibits a spin-polarized triplet EI state.

\paragraph*{Exciton dispersion.} 
The $\mathbf{q}$-dependent lowest $E_t$ of Ta$_3X_8$ monolayers along the high-symmetry paths is shown in Fig.~\ref{fig-EI-q}(b). 
The $E_t$ remains negative across all $\mathbf{q}$ values, which indicates that the EI state is accessible for all momenta in the full BZ.
Although Ta$_3X_8$ monolayers host an indirect $E_g$, the exciton with $\mathbf{q}=0$ is energetically more stable than those with $\mathbf{q}\neq0$.
This indirect-to-direct transition crossover between the single-particle band and the exciton dispersion has also been predicted in semihydrogenated graphene and may be linked to nonlocal dielectric screening~\cite{TripletFM-PhysRevLett.124.166401}. 
In addition, Ta$_3X_8$ monolayers exhibit small exciton dispersion with bandwidths narrower than 13 meV, a direct consequence of their electronic flat bands and spin-triplet nature~\cite{EIband-PhysRevLett.116.066803}.
The momentum-dependent dispersion of exciton may be experimentally probed using momentum-resolved electron energy loss spectroscopy or resonant inelastic X-ray spectroscopy~\cite{EIband-PhysRevLett.116.066803,TripletFM-PhysRevLett.124.166401,energyloss-Egerton2011,x-ray-Schülke2007,x-ray-RevModPhys.83.705,WSe2-TEIwave-PhysRevLett.134.066602,CrI3-PhysRevX.15.011042}.
Recently, the dispersion of two dark excitons in vdW FM CrI$_3$, exhibiting narrow bandwidths ($\sim$10 meV), has been experimentally observed using high-resolution resonant inelastic X-ray scattering~\cite{CrI3-PhysRevX.15.011042}. 
These two dark excitons also originate from the spin-flipping transitions and may strongly interact with magnons as they propagate through the lattice~\cite{CrI3-PhysRevX.15.011042,CrI3-topspin-PhysRevX.8.041028,CrI3-excitonic-NC2019}.

\paragraph*{Conclusion and discussion.}
Since excitons are charge-neutral quasi-particles, their uniform flow does not generate an electrical current, making it challenging to detect exciton condensation through traditional transport measurements. 
It is worth noting that spin-polarized triplet excitons have $ S_z =1$.
Therefore, the BEC of these spin-polarized triplet excitons can lead to the emergence of a spin supercurrent~\cite{spinsuperconductor-PhysRevB.84.214501,spinsuperconductor2-ABC-PhysRevB.86.085441,spinsuperconductor3-ABCA-PhysRevB.110.174512,TripletFM-PhysRevLett.124.166401}, which can be detected through magnetic transport experiments~\cite{spincurrent-doi:10.1126/sciadv.aat1098}. 
In addition to exhibiting zero spin resistance, the spin-polarized triplet exciton condensate can exhibit an electric “Meissner effect” against a spatially varying electric field~\cite{spinsuperconductor-PhysRevB.84.214501}, and thus is referred to as a spin superconductor.
Unlike parabolic bands, the low-energy flat bands can make excitons condense into an ideal form like one-body bosons and improve exciton coherency~\cite{flatEI2-PhysRevLett.130.186401}.
The BEC of spin-polarized triplet excitons is as physically significant as the spin-triplet superconductivity~\cite{Spin-triplet-superPhysRevB.100.035203,Spin-triplet-superdoi:10.1126/science.aav8645} and the superfluidity of $^3$He~\cite{He3-RevModPhys.47.331,He3AnnualReviews}.
More importantly, due to potential magnetoelectric coupling~\cite{ME-nature05023,ME-Spaldin2019,Ta3I8-acs.jpclett.4c00858,Ta3I8-APL,Ti3X8-FMFE-PhysRevB.104.L060405}, the magnetic moments may be reversed through a breathing process in the Kagome lattice triggered by an applied vertical electric field. 
If so, two spin supercurrent states with $S_z=1$ and $S_z=-1$ can be switched by an electric field~\cite{AMFE-spintriplet2023}, which is attractive for potential applications in spintronic devices, such as spin Josephson junctions~\cite{Josephson-Senapati2011,Josephson-PhysRevLett.108.127002,spinsuperconductor-PhysRevB.84.214501}.

The spin-polarized triplet EI discussed in this work differs from the TR invariant spin-triplet EI recently proposed in nonmagnetic diatomic Kagome lattices~\cite{flatEI-PhysRevLett.126.196403} and quantum spin-Hall insulators (AsO and Mo$_2$TiC$_2$O$_2$)~\cite{parity-tp-triplet-PhysRevB.109.075167-2024}.
In the TR invariant spin-triplet EI, the exciton states with $S_z=1$ and $S_z=-1$ are degenerate.
The spontaneous BEC of these degenerate excitons does not form a spin superconductor state and cannot support a spin supercurrent.

In summary, we predict that Ta$_3X_8$ FM monolayers are spin-polarized triplet EIs, as revealed through systematic first-principles $GW$+BSE calculations.
Single-particle calculations reveal that these monolayers are intrinsic BMSs.
The electron-hole transitions between the two low-energy flat bands are parity- and spin-forbidden, which effectively inhibits dielectric screening and enables the EI state.
The calculated $E_b$ of the lowest-energy exciton with all $\mathbf{q}$ exceeds the $E_g$, indicating that Ta$_3X_8$ monolayers belong to a many-body EI ground state characterized by spontaneous exciton BEC.
Analysis of the exciton wavefunction further reveals that the lowest-energy exciton is a tightly bound Frenkel-like state, exhibiting a spin-polarized triplet nature with $ S_z =1$. 
This study presents a robust material platform that not only enables fundamental exploration of spin-polarized triplet EIs but also paves the way for developing next-generation spintronic devices with enhanced quantum efficiency and spin manipulation capabilities.

\ \\
\paragraph*{Acknowledgments.}
This work was supported by the National Natural Science Foundation of China (Grants No. 12188101), National Key R\&D Program of China (Grants No. 2022YFA1403800), and the Center for Materials Genome.

%\nolinenumbers

%\bibliography{refs}
%merlin.mbs apsrev4-1.bst 2010-07-25 4.21a (PWD, AO, DPC) hacked
%Control: key (0)
%Control: author (8) initials jnrlst
%Control: editor formatted (1) identically to author
%Control: production of article title (-1) disabled
%Control: page (0) single
%Control: year (1) truncated
%Control: production of eprint (0) enabled
%

%\ \\

\clearpage

\end{document}